\documentclass[fleqn,usenatbib]{mnras}

\usepackage{newtxtext,newtxmath}

\usepackage[T1]{fontenc}
\usepackage{ae,aecompl}

\usepackage{graphicx}	
\usepackage{amsmath}	
\usepackage{amssymb}	

\newcommand{\kms}{\,km\,s$^{-1}$} 
\newcommand{\kmsmpc}{\,km\,s$^{-1}$Mpc$^{-1}$}
\newcommand{\beq}{\begin{equation}}
\newcommand{\eeq}{\end{equation}}
\newcommand{\bea}{\begin{eqnarray}}
\newcommand{\eea}{\end{eqnarray}}
\newcommand{\lsim}{\mbox{$\:\stackrel{<}{_{\sim}}\:$} }
\newcommand{\gsim}{\mbox{$\:\stackrel{>}{_{\sim}}\:$} }
\newcommand{\om}{\rm \Omega_{\rm m}}
\newcommand{\oll}{\rm \Omega_\Lambda}

\newcommand{\zbar}{\bar{z}}

\title[Hubble constant -- redshift effects]{Can redshift errors bias measurements of the Hubble Constant?}

\author[Davis, Hinton, Howlett, \& Calcino]{Tamara M.\ Davis$^{1}$\thanks{Contact e-mail: \href{mailto:tamarad@physics.uq.edu.au}{tamarad@physics.uq.edu.au}}\thanks{Present address: School of Mathematics and Physics, The University of Queensland. QLD 4072, Australia.}
Samuel R.\ Hinton,$^{1}$
Cullan Howlett,$^{1}$ and
Josh Calcino$^{1}$\\
$^{1}$School of Mathematics and Physics, The University of Queensland. QLD 4072, Australia.}

\date{Accepted 2019 September 8. Received 2019 September 6; in original form 2019 July 26}

\pubyear{2019}

\begin{document}
\label{firstpage}
\pagerange{\pageref{firstpage}--\pageref{lastpage}}
\maketitle

\begin{abstract}
Redshifts have been so easy to measure for so long that we tend to neglect the fact that they too have uncertainties and are susceptible to systematic error. As we strive to measure cosmological parameters to better than 1\% it is worth reviewing the accuracy of our redshift measurements. Surprisingly small systematic redshift errors, as low as $10^{-4}$, can have a significant impact on the cosmological parameters we infer, such as $H_0$. 
Here we investigate an extensive (but not exhaustive) list of ways in which redshift estimation can go systematically astray. We review common {\it theoretical} errors, such as adding redshifts instead of multiplying by $(1+z)$; using $v=cz$; and using only cosmological redshift in the estimates of luminosity and angular-diameter distances. We consider potential {\it observational} errors, such as rest wavelength precision, air to vacuum conversion, and spectrograph wavelength calibration. Finally, we explore {\it physical effects}, such as peculiar velocity corrections, galaxy internal velocities, gravitational redshifts, and overcorrecting within a bulk flow. 
We conclude that it would be quite easy for small systematic redshift errors to have infiltrated our data and be impacting our cosmological results. While it is unlikely that these errors are large enough to resolve the current $H_0$ tension, it remains possible, and redshift accuracy may become a limiting factor in near future experiments. With the enormous efforts going into calibrating the vertical axis of our plots (standard candles, rulers, clocks, and sirens) we argue that it is now worth paying a little more attention to the horizontal axis (redshifts).
\end{abstract}

\begin{keywords}
cosmology: theory -- galaxies: distances and redshifts -- methods: observational
\end{keywords}



\section{Introduction}

The Hubble Constant, $H_0$, quantifies the current expansion rate of the Universe.  It is among the most important cosmological parameters not only because it determines many of the important cosmological features of our universe (such as age), but also because improved knowledge of $H_0$ allows one to improve the precision of the other cosmological parameters (such as the matter density and dark energy density). 

\subsection{Local vs Global}
Enormous effort has gone into measuring $H_0$ over the last century.  Modern measurements focus on two main techniques.  Firstly, ``local'' measurements, which use a distance ladder built up from nearby distance measures and reaching to redshifts of approximately $z\sim0.40$.  Secondly, ``global'' measurements, which fit $H_0$ simultaneously with other cosmological parameters to large scale features of the universe such as the cosmic microwave background (CMB) or baryon acoustic oscillations (BAO).  

Local measurements include techniques such as the distance ladder out to type Ia supernovae calibrated through geometric distances $+$ Cepheid variables, or by the tip of the red giant branch (TRGB); and gravitational wave standard siren measurements \citep[e.g.][]{riess19,freedman19,abbott17}.

Global measurements include measurements of large scale structure; the BAO standard ruler; the CMB standard ruler, and weak lensing \citep[e.g.][]{boss17,planck18,des18cosmo}.  These are called global in part because they are on much larger scales, but primarily because you do not measure $H_0$ directly -- instead you fit an entire cosmological model, in which $H_0$ is one of the parameters.

Gravitational lensing time-delay $H_0$ measurements \citep[e.g.][]{wong19} fall somewhere in between, found at moderate redshifts they usually do need knowledge of the cosmological model \citep[with some exceptions e.g.][]{paraficz10}, similar to global measurements.  They use a combination of distances that cumulatively make them more akin to a standard ruler, but since they are single systems they also share some aspects of the local measurements.

Another option includes anchoring a distance ladder measurement to high-redshift calibrators, instead of low-redshift ones.  For example, one can anchor the supernova Hubble diagram to high-redshift BAO distance measurements \citep{aubourg15,macaulay19} or lensing time delay measurements \citep{wojtak19}, instead of local distance ladder measurements.  This is known as the ``inverse distance ladder'' technique.  

\subsection{Candles vs Rulers}
Another way to classify the types of $H_0$ measurements is to consider the difference between standard candles and standard rulers. The local techniques tend to be standard candles (Cepheids, TRGB, supernovae, and gravitational waves), while the global tend to use standard rulers (BAO, CMB, and gravitational lensing time delays). 
In other words, local measurements rely on luminosity distances, while global rely on angular diameter distances.  

It is possible that the current controversy over the value of $H_0$ is rooted in a difference between standard candles (luminosity distance, $D_L$) and standard rulers (angular diameter distance, $D_A$).  Luminosity distance and angular diameter distance are famously related by $D_L = D_A (1+z)^2$, where $z$ is redshift.  
This formula encompasses the principle of distance duality, which states that luminosity distances and angular diameter distances must always be related by exactly a factor of $(1+z)^2$ in all metric theories of gravity \citep{etherington33}.  However, if you use an incorrect redshift in the calculations then distance duality may appear to be broken.

This difference inspired us to consider the impact of a systematic error in our redshift measurements.  It is possible that since luminosity distances involve multiplying the curvature-corrected comoving distance by $(1+z)$ while angular diameter distances involve dividing by $(1+z)$, that an error in the redshift would bias the distances in opposite ways, biasing our measurement of $H_0$.

Other papers have looked at the impact of statistical uncertainties in redshifts \citep[e.g.][]{huterer04,aldering07,chaves18}, but here we concentrate on systematic uncertainties.  


We showed in \citet[][Fig.~6]{wojtak15} that a redshift bias of $10^{-4}$ could bias our inference of the equation of state of dark energy, $w$, by 0.05.  Then in \cite{calcino17} we showed that a $5\times10^{-4}$ redshift bias could cause a 1~\kmsmpc\ bias in $H_0$.   \citet{linder19} showed that for photometric redshifts used in supernova studies systematics need to be controlled at the level of a few $\times10^{-3}$.  Thus a surprisingly small redshift shift can have a significant impact on our cosmological inferences.  Given the sensitivity of $H_0$ to small redshift biases, it is worth checking our current data carefully before resorting to more exotic explanations of any $H_0$ tensions.  It also makes it clear that to avoid biases in future cosmological measurements aiming for better than 1\% precision we need to pay renewed attention to redshift calibration.

This paper is structured as follows.  Firstly, we define the mathematical foundation of $H_0$ measurements.  Even though most of this appears in standard text books, there are some subtleties involved that make it worth outlining carefully, such as which redshifts should be used in different parts of each equation, and which approximations are appropriate.  We outline where some standard tools such as the NASA Extragalactic Database (NED, \url{https://ned.ipac.caltech.edu/}) make invalid approximations.  Secondly, we explore in detail the size of a redshift bias that would be needed to explain the current discrepancy in $H_0$ measurements.  Then we assess whether such a bias could be in our data.

\section{Mathematical foundations}\label{sect:math}

Starting from the metric,\footnote{Where $c$ is the speed of light, $dt$ is the time separation, $d\chi$ is the comoving coordinate separation and $d\psi^2=d\theta^2+\sin^2\theta d\phi^2$, where $\theta$ and $\phi$ are the polar and azimuthal angles in spherical coordinates. The scalefactor, $R$, has dimensions of distance.   The function $S_k(\chi)=\sin\chi$, $\chi$ or $\sinh\chi$ for closed ($k=+1$), flat ($k=0$), or open ($k=-1$) universes respectively. }
\beq  
ds^2 = -c^2dt^2 + R(t)^2[d\chi^2+S_k^2(\chi)d\psi^2], 
\label{eq:frwmetric}
\eeq
we see the radial ($d\psi=0$) distance along a constant time-slice ($dt=0$) is $ds = R d\chi$, which upon integrating gives the proper distance,
\beq D=R\chi .\eeq
Differentiating with respect to $t$, assuming comoving distance is constant ($d\chi/dt=0$), gives the Hubble-Lema\^itre law, 
\beq v = \frac{dD}{dt} =\frac{dR}{dt} \chi = \frac{\dot{R}}{R}R\chi = HD ,\eeq
where $\dot{R}\equiv dR/dt$ and the standard definition of the Hubble parameter is,
\beq H \equiv \frac{\dot{R}}{R}.\eeq 
We define the normalised scalefactor $a\equiv R/R_0$, where subscript 0 denotes quantities at the present day.  Thus $a_0=1$, from which one can see that the $H$ at the present day is $H_0=da/dt$, which is the current normalised rate of expansion of the universe.  However, in general it is incorrect to call $H$ the expansion rate of the universe, it is the expansion rate relative to the scalefactor of the universe at that time.  Even in an accelerating universe $H$ tends to decrease over time because the denominator (scalefactor) increases more quickly than the numerator (expansion rate), unless the universe is accelerating at a faster than exponential rate. 

Now consider the path of a photon ($ds=0$), 
\beq c dt = R(t) d\chi. \eeq
This equation shows that the velocity of light is purely a peculiar velocity $c=R\,d\chi/dt$, independent of the expansion of the universe -- which is one reason the recession velocity in the Hubble-Lema\^{i}tre law is not bounded by the speed of light \citep{davis04}.\footnote{Note, this is a coordinate dependent statement, and arises in part because we have chosen a time slice, $t$, such that the universe is homogeneous.  From a special relativistic perspective this is a strange time coordinate, since observers moving relative to each other (comoving observers) agree on the rate of clocks ticking.}
To calculate the comoving coordinate $\chi$ between two points on the photon's path you need to integrate
\beq \int_{t_{\rm e}}^{t_{\rm o}}d\chi = c\int_{t_{\rm e}}^{t_{\rm o}} \frac{dt}{R(t)},\label{eq:dchi} \eeq
where $t_{\rm e}$ is the time the photon was emitted, and $t_{\rm o}$ the time it was observed (we do not have to be the observers, $t_{\rm o}$ could be anytime).   
 
 Substituting $1+z=R_0/R$ and integrating from now back to the time a photon at cosmological redshift $\zbar$ was emitted we get,
 \bea \chi(\zbar) &=&\frac{c}{R_0}\int_0^{\zbar}\frac{dz}{H(z)},\label{eq:chi}\\
                            &=&\frac{c}{R_0H_0}\int_0^{\zbar}\frac{dz}{E(z)}, \label{eq:Ez}
\eea
where $E(z)\equiv H(z)/H_0$ is independent of $H_0$.\footnote{Here $E(z)=\left[\sum_i\Omega_i a^{-3(1+w_i)}\right]^{1/2}$ where $\Omega_i$ are normalised densities of components of the universe (matter, radiation, and dark energy) each with a different equation of state $w_i$.}  Since the dimensionless normalised curvature $\Omega_k$ follows $|\Omega_k|^{1/2}=c/(R_0H_0)$, you can see that the comoving coordinate as a function of redshift is independent of both $H_0$ and $R_0$.

\subsection{Types of distance} 
The comoving distance ($R_0\chi$) is arguably the most fundamental distance, but it is not typically what we measure.   For convenience we define the `curvature corrected distance' as,
 \beq \tilde{D} = R S_k(\chi), \label{eq:curvedist}\eeq
 which becomes useful for any measurements involving distances perpendicular to the line of sight -- such as the surface area of a sphere propagating from a light source.
 
 The luminosity distance is the distance that makes the standard inverse radius squared law true, i.e.\ ${\rm Flux} = {\rm Luminosity}/(4\pi D_{\rm L}^2)$.  In an expanding universe the flux is decreased both by time dilation reducing the rate of photon arrival and by redshift decreasing the energy of each photon, such that 
 \beq D_L = \tilde{D}(1+z). \label{eq:DL}\eeq
In contrast the angular diameter distance is the distance that makes the standard angular distance equation\footnote{Note that for large angles as seen for nearby BAO or large modes in the CMB, this small angle approximation is not sufficient and one has to take into account the difference between a chord and an arc.}  true, $D_A = r/\theta$ for some perpendicular ruler $r$ subtending an angle $\theta$.  Because it is the apparent size at the time of emission that matters, a $(1+z)$ factor enters making
\beq D_A = \tilde{D}/(1+z). \label{eq:DA}\eeq

There is an important difference in the redshifts that appear in Eq.~\ref{eq:chi} and those that appear in Eqs.~\ref{eq:DL}, \ref{eq:DA}.  The redshift that determines the comoving coordinate is the cosmological redshift.  That is, the redshift purely due to the expansion of the universe.  It is the redshift that would be measured in the CMB frame if neither the emitter nor receiver had any peculiar velocity or gravitational redshifts.  In contrast, the features that cause the $(1+z)$ factors in the luminosity and angular diameter distances are features that arise due to the total redshift, including peculiar velocities and gravitational redshifts.  For example, the time dilation factor of $(1+z)$ in luminosity distance cares about the actual time dilation experienced, which includes effects of peculiar velocity and gravitational redshift -- so if you remove them and use the cosmological redshift you have inappropriately removed a real effect.  Therefore the luminosity and angular diameters distances should be more carefully written as
\bea D_L(\zbar,z_{\rm obs}) &=& \tilde{D}(\zbar)(1+z_{\rm obs}), \\
     D_A(\zbar,z_{\rm obs}) &=& \tilde{D}(\zbar)/(1+z_{\rm obs}). \eea
The difference is negligible for most situations \citep{calcino17}, but can become significant for high source peculiar velocities and can be systematic if all emitters or receivers are in a gravitational well.  Therefore it may be important for very precise studies.  Also, making observational errors in $z_{\rm obs}$ could influence luminosity and angular diameter distances in different directions, and therefore is worth investigating in the context of the $H_0$ controversy. 

\subsection{Deriving $H_0$ from observables}

Having established how the various distance measures relate to the Hubble-Lema\^{i}tre Law, let us turn to how the value of the $H_0$ is measured in practice using standard candles.  
The heart of the method is to use a measured redshift, $z$, to infer a velocity, $v(z)$, then use a measured magnitude to infer a distance, $D(z)$.  Then  take the ratio of the two to infer the Hubble constant, 
\beq H_0 = v(z)/D(z), \label{eq:H0}\eeq
where $v(z)$ and $D(z)$ are the present day velocity and proper distance of a comoving galaxy at redshift $z$.  Averaging over many such measurements gives the same answer as fitting a line to find the $a_x$ parameter used in \citet{riess16}; see Appendix~\ref{app:riess}.

Since $v(z)\equiv H_0 D = H_0R_0\chi(z)$ it is easy to see from Eq.~\ref{eq:Ez} that you can derive velocity directly from redshift, independently of $H_0$,
\beq v(z) = c\int_0^{\bar{z}}\frac{dz}{E(z)}. \label{eq:correctvz}\eeq 
$E(z)$, and thus $v(z)$, depend on the cosmological model.  
At very low redshift ($z\lsim 0.01$) the cosmological dependence is small and velocity can be approximated by $v=c\bar{z}$, 
but at higher redshifts it rapidly becomes inaccurate. 
To take this into account while trying to mitigate the cosmological-model dependence of $v(z)$ a common approximation is used, which is expressed in terms of deceleration parameter, $q_0$, and jerk $j_0$,
\beq v(z) = \frac{c z}{1+z} \left[1+\frac{1}{2}(1-q_0)z - \frac{1}{6}(1-q_0-3q_0^2+j_0)z^2\right], \label{eq:vzHD}\eeq
with $q_0=-0.55$ and $j_0=1.0$ for standard $\Lambda$CDM with $(\om,\oll)\sim(0.3,0.7)$, \citep{riess09,riess11,riess16}. 
While it is often claimed this makes the result independent of cosmology, that is not entirely true -- if you used the CDM model $(\om,\Omega_{\Lambda})=(1.0,0.0)$ parameters for this expansion, which are $q_0=0.5$ and $j_0=1.0$ then you would get a $-3.5$\kmsmpc\ error in $H_0$ at $z\sim0.1$.   Nevertheless, within the observationally favoured range of parameters for our standard model of the universe this remains an adequate approximation out to about $z\sim0.3$ (see Fig.~\ref{fig:H0error_vs_z} for an example of how it fails at higher $z$).  There is no real way to avoid the cosmological depenence of $v(z)$, even though the cosmological dependence may be weak. For the most accurate results for $H_0$ we should actually just use the exact formula, Eq.~\ref{eq:correctvz}, for the best fitting cosmological model, or marginalise over this uncertainty \citep[e.g.][]{marra13}. 
 
\subsection{Combining redshifts}\label{sect:combiningz}
Finally, to conclude the mathematical basis summary, we consider how different contributions to redshifts are combined.  
Redshift is defined as the difference in wavelength divided by the emitted wavelength, $z=(\lambda_{\rm o}-\lambda_{\rm e})/\lambda_{\rm e}$, where subscripts `o' and `e' mean observed and emitted respectively.   Thus $1+z = \lambda_{\rm o}/\lambda_{\rm e}$.    Imagine we are observing a comoving object, which emitted light at $\lambda_{\rm 0}$ that arrived at our position with $\lambda_{\rm 1}$.  Locally, we have a peculiar velocity which causes an additional redshift, where the the wavelength we should have seen $\lambda_{\rm 1}$ is shifted to $\lambda_{\rm 2}$.  Thus the redshift we observe is given by, 
\beq 1+z_{\rm 02} = \frac{\lambda_2}{\lambda_0} = \frac{\lambda_2}{\lambda_1}\frac{\lambda_1}{\lambda_0} = (1+z_{\rm 21})(1+z_{\rm 10}). \eeq
In practice, the observed redshift includes many components that are not cosmological, including the redshift due to Earth's peculiar motion around the Sun ($z_{\rm p}^{\rm Earth}$), the Sun's peculiar motion with respect to the comoving rest frame of the universe, i.e.\ with respect to the CMB ($z_{\rm p}^{\rm Sun}$), the peculiar motion of the emitting object ($z_{\rm p}^{\rm source}$), and the gravitational redshifts experienced along the light's path ($z_{\rm grav}$).   Most cosmological measurements need to be done in the comoving frame; using the cosmological redshift ($\bar{z}$).   The redshift we observe is given by, 
\beq 1+z_{\rm obs} = (1+\bar{z})(1+z_{\rm p}^{\rm Earth})(1+z_{\rm p}^{\rm Sun})(1+z_{\rm p}^{\rm source})(1+z_{\rm grav}). \label{eq:zs}\eeq
Often you will instead see approximate equations such as, 
\beq z_{\rm obs} = \bar{z} + z_{\rm p}^{\rm Earth} + z_{\rm p}^{\rm Sun} + z_{\rm p}^{\rm source} + z_{\rm grav}.\eeq 
 From the above derivation you can see this is only true to first order, yet this approximation still frequently appears, for example in the NASA Extragalactic Database (NED),\footnote{\url{https://ned.ipac.caltech.edu/help/zdef.html}} and we explore the impact of using the approximation in Sect.~\ref{sect:zcorr}.

A similar error occurs in peculiar velocity calculations when using $v=cz_{\rm obs}$ for the total velocity.  Many peculiar velocity papers start with the incorrect equation, 
\beq v_{\rm p} = cz_{\rm obs}-H_0D,\eeq 
which is a rearrangement of the correct equation, 
\beq v_{\rm total}=v_{\rm r}+v_{\rm p}\eeq 
with recession velocity correctly given by $v_{\rm r}=H_0D$, but with the total velocity erroneously given by $v_{\rm total}=cz_{\rm obs}$.   

Again, this approximation appears in the NED velocity conversion calculator notes (i.e.\ using $v=cz$ for the velocity in the Hubble-Lema\^{i}tre law; $D_P=cz/H_0$ in their notation), and is often seen in peculiar velocity papers.  However, it performs badly beyond $z\lsim0.01$, and by $z\sim0.1$ it gives an error of 700\kms, which is about twice the typical peculiar velocity of galaxies.  We discussed the impact of this error in \citet{davisscrim14}.

\section{Small redshift biases matter}\label{sect:zbias}
Redshifts have been so easy to measure for so long that they usually do not even get error bars when plotted on graphs.  The most precise redshifts currently achievable (as used for planet searches or time-varying fundamental constant searches) can reach a  precision of better than $\sim 10^{-7}$ \citep{lovis07}.   However, a typical large redshift survey such as SDSS at the Apache Point Observatory or OzDES on the Anglo-Australian Telescope quotes their redshifts to a precision of $\sim 10^{-3}$.  

\begin{figure}
    \centering
    \includegraphics[width=84mm]{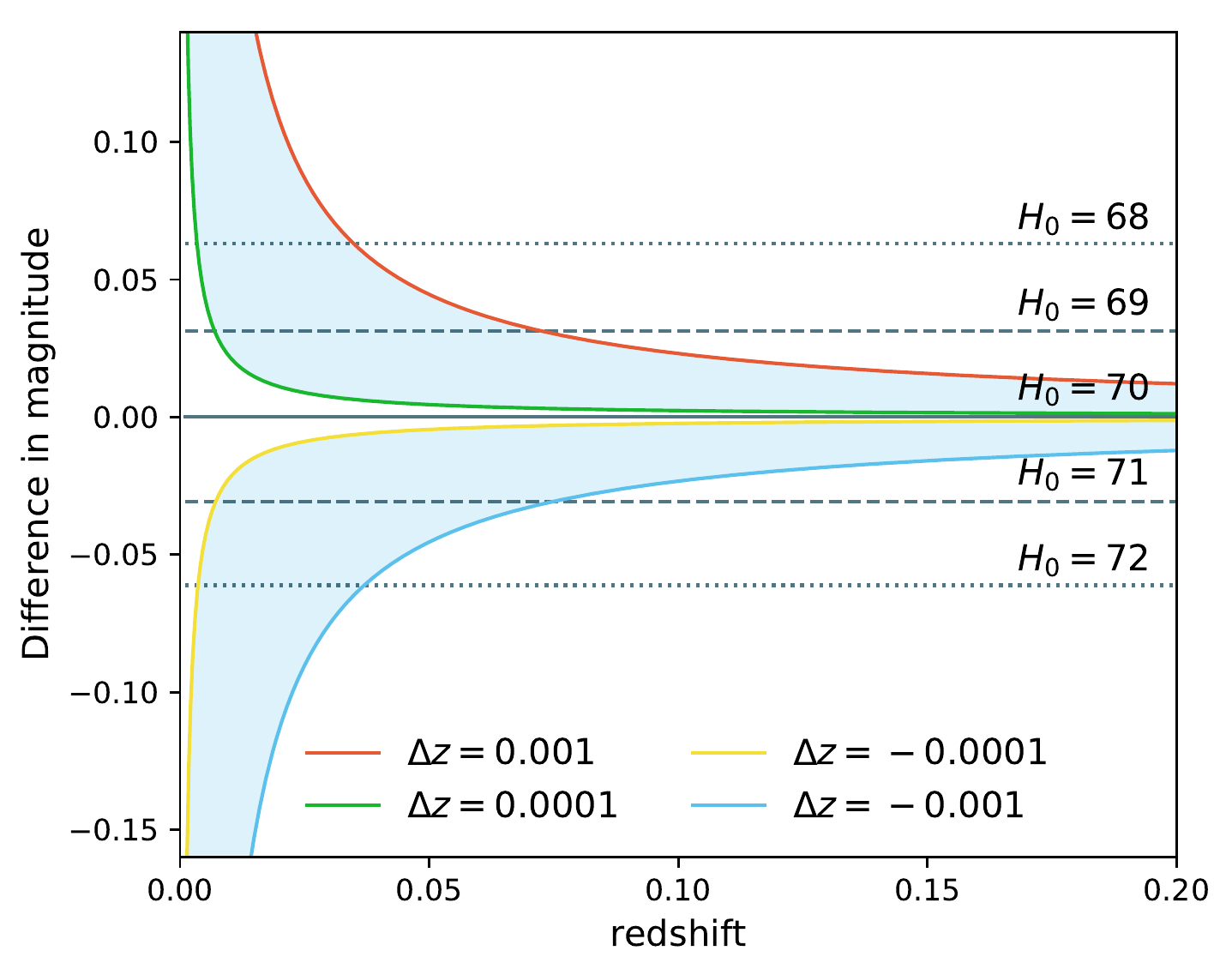}
    \caption{The impact on the Hubble diagram of changing $H_0$, compared to the difference induced by an error in redshift.  In each case we show apparent magnitude for the model in question, minus the apparent magnitude for a fiducial model with $H_0=70$\kmsmpc\ and no redshift error.  All $H_0$ measurements are in \kmsmpc, and $(\om,\oll,w)=(0.3,0.7,-1.0)$ in each case.  Compare to Fig.~4 in \citet{wojtak15}.}
    \label{fig:magdiff}
\end{figure}

So can a redshift error of $<10^{-3}$ significantly bias our cosmological measurements?  In Figure 4 of \citet{wojtak15} we show that the shift in magnitude  caused by a 1\% variation in $\om$, $\oll$, or $w$, is similar to the change in magnitude caused by a redshift error on the order of 10$^{-5}$ to $10^{-4}$.  So our redshifts have to be unbiased at the level of $10^{-4}$ if we want cosmological inferences from the Hubble diagram to be unbiased at the level of $1\%$.  For that analysis we only considered relative magnitudes, as usually done for supernova dark energy measurements when we marginalise over absolute magnitude and $H_0$.   In Figure~\ref{fig:magdiff} above we repeat that analysis for cosmological models with different $H_0$.  When measuring $H_0$ we do not marginalise over absolute magnitudes, so there is less flexibility in the fit.  Figure~\ref{fig:magdiff} shows that a systematic redshift bias of between $10^{-4}$ and $10^{-3}$ at redshifts less than $\sim0.1$ can give a similar shift in magnitude to a $H_0$ change of about 1~\kmsmpc, and the shift rapidly increases as you go to lower redshifts.  While not enough to fix the Hubble constant tension on its own, it is enough to cause a significant change in the inferred $H_0$. 

We assessed the impact of a redshift bias on both $\om$ and $H_0$ in \cite{calcino17}, and found that redshift accuracy of $10^{-4}$ was needed for unbiased cosmological measurements given current precision.  We also found that allowing a redshift bias as a free parameter in the cosmological fit to type Ia supernovae naturally aligned the $\om$ measurements from supernovae and CMB, but the uncertainty on the best fit redshift bias was as large as the best fit bias itself.   We examine the effect of redshift bias on $H_0$ in more detail here.  

Some redshift systematic errors would enter as a multiplicative factor of $(1+\Delta z)$, particularly when due to physical effects.  However, we look at a simple additive error, where we think we have measured a true $z$ but we have actually measured a biased $z+\Delta z$, where $\Delta z$ is the size of the systematic error.  This is appropriate for most measurement errors. 

\begin{figure}
    \centering
    \includegraphics[width=82mm]{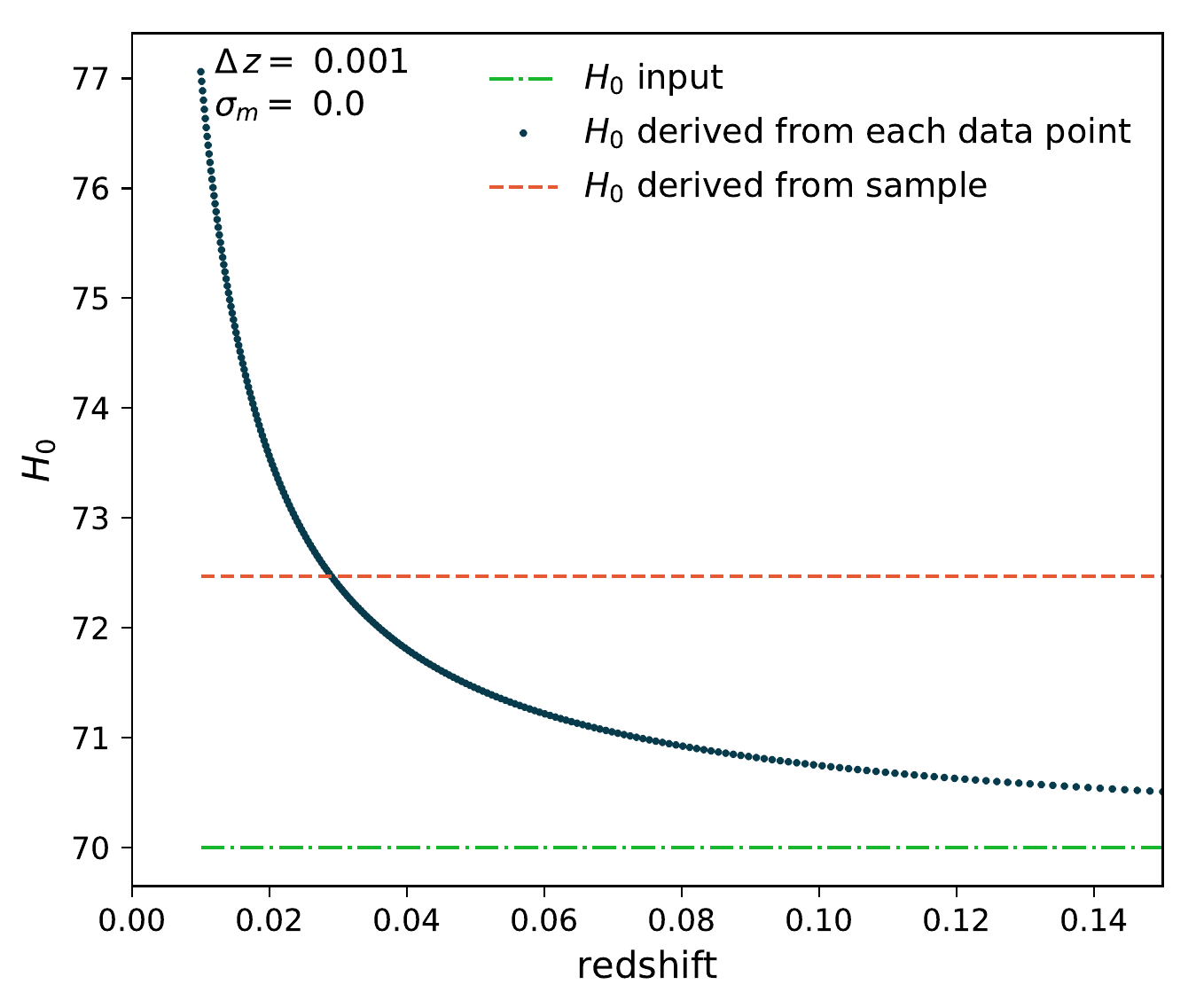}\vspace{-2mm}
    \includegraphics[width=82mm]{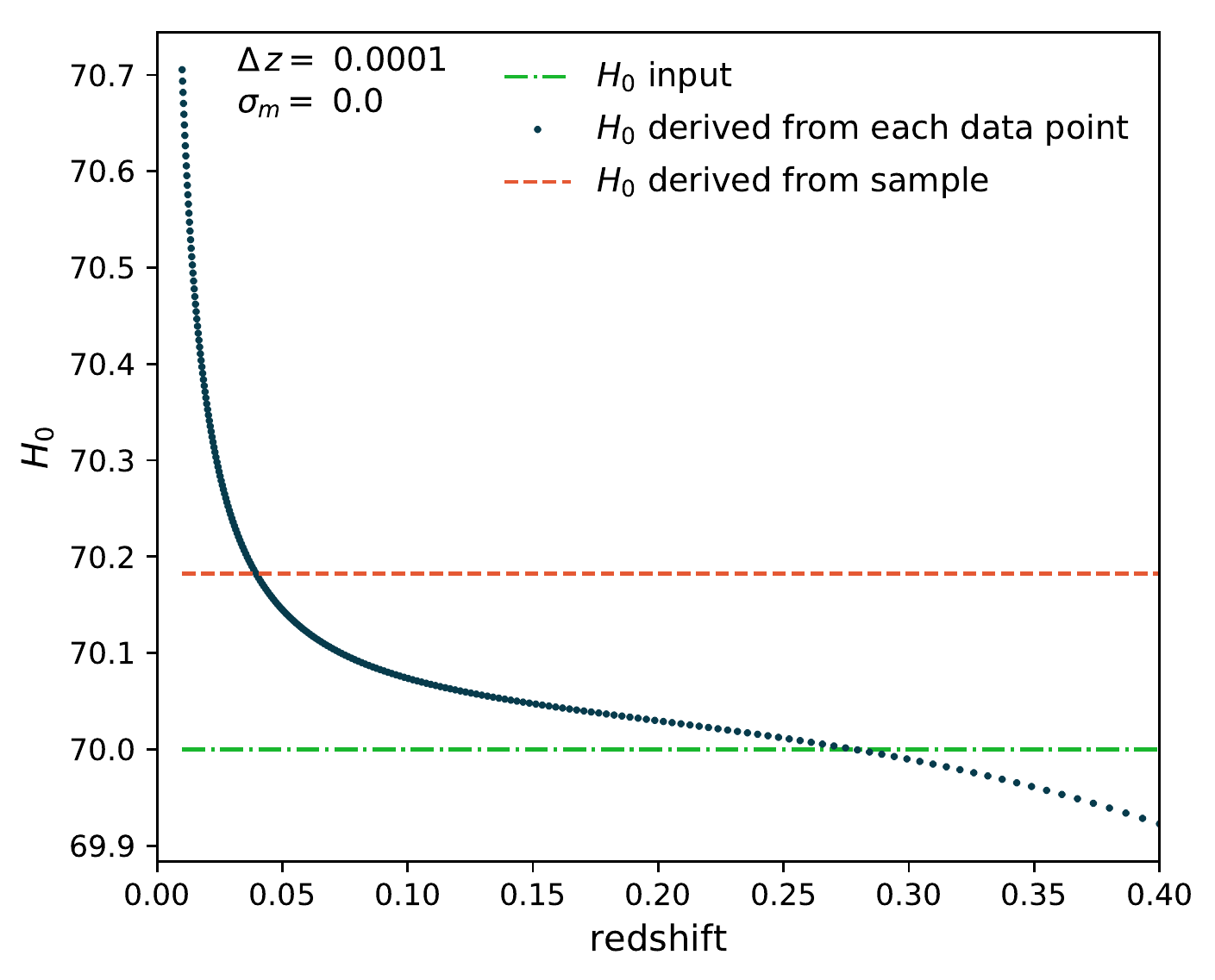}
    \caption{The dark blue points show the error in $H_0$ as a function of redshift for a redshift bias of $10^{-3}$ (upper) and $10^{-4}$ (lower).  The horizontal green dot-dashed line shows the input $H_0$ and the red dashed line shows the resulting $H_0$ measurement for the sample.  The upper plot has a maximum redshift of 0.15, while the lower plot extends to 0.40, both have a minimum redshift of 0.01.  At low redshifts the deviation is due to the redshift bias.  At higher redshifts a deviation is seen due to the use of the $v(z)$ approximation of Eq.~\ref{eq:vzHD}.  } \vspace{-3mm}
    \label{fig:H0error_vs_z}
\end{figure}

\begin{figure}
    \centering
    \includegraphics[width=82mm]{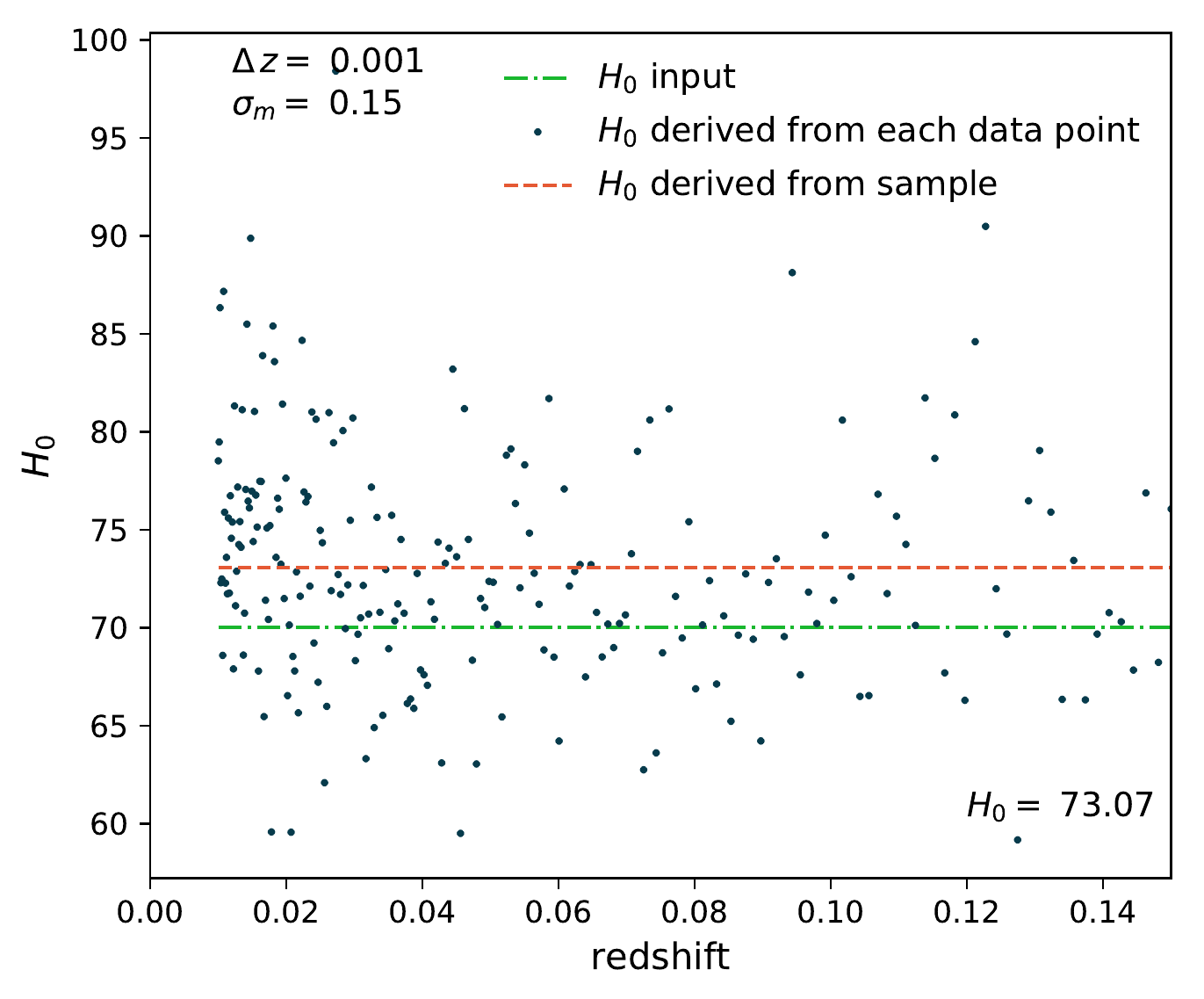}
    \caption{As for the upper panel of Fig.~\ref{fig:H0error_vs_z}, but with Gaussian random scatter of $\sigma_m=0.15$ added to the data.  The scatter is larger than the signal of the bias, so it would be difficult to diagnose a redshift bias just given this data, even the extreme bias of $\Delta z = 10^{-3}$ shown here.  This particular realisation results in an $H_0$ measurement of 73.07~\kmsmpc\ when the true $H_0$ is 70~\kmsmpc.  The result will be dependent on the realisation and the redshift distribution (in this case $0.01<z<0.15$ spread logarithmically).}
    \label{fig:H0error_vs_z_random}
\end{figure}

In Fig.~\ref{fig:H0error_vs_z} we show two plots of the $H_0$ we would infer from a set of standard candles with a redshift bias of $10^{-3}$ (upper) or $10^{-4}$ (lower), over two different redshift ranges.  Each dark blue dot represents a standard candle measurement at a particular redshift.  We generate fake magnitude-redshift data of $\sim200$ standard candles, spread evenly across the redshift range in log space, using a fiducial model of $H_0=70$\kmsmpc\ and $(\om,\oll,w)=(0.3,0.7,-1.0)$, apply an additive redshift offset $\Delta z$, then measure $H_0$ using a variant of Eq.~\ref{eq:H0} adapted to luminosity distance.  That is, we use $H_0=v(z)(1+z)/D_L(z)$ with the luminosity distance measured correctly but using the $v(z)$ approximation of Eq.~\ref{eq:vzHD} and both terms in the numerator using the biased redshift $\bar{z}+\Delta z$.  Strong deviations are seen at low redshift due to $\Delta z$.  The smaller high-redshift deviation in the lower panel arises because of the $v(z)$ approximation failing (Eq.~\ref{eq:vzHD}).  

The dashed red line shows the resulting $H_0$ measurement from averaging all the $H_0$ values in the sample.  (We checked that this gives the same result as fitting for the $a_x$ parameter used in \citet{riess16} as described in Appendix~\ref{app:riess}.) 

The deviations from a horizontal line in Fig.~\ref{fig:H0error_vs_z} are so drastic that it would seem that such a bias must be obvious if it was in the data.  However, this data was generated to be perfect, without uncertainties.  Adding a magnitude uncertainty of $\sigma_m=0.15$ results in Fig.~\ref{fig:H0error_vs_z_random}, where a Gaussian random offset with $\sigma_m=0.15$ has been applied to the data points.  The dispersion is much higher than the signal, which makes it unlikely we would have seen noticed such a redshift bias in our data. 

To explore how this could impact existing supernova measurements we reproduce some of the plots from \citet{riess16} in Appendix~\ref{app:riess}.

\begin{figure}
    \centering
    \includegraphics[width=84mm]{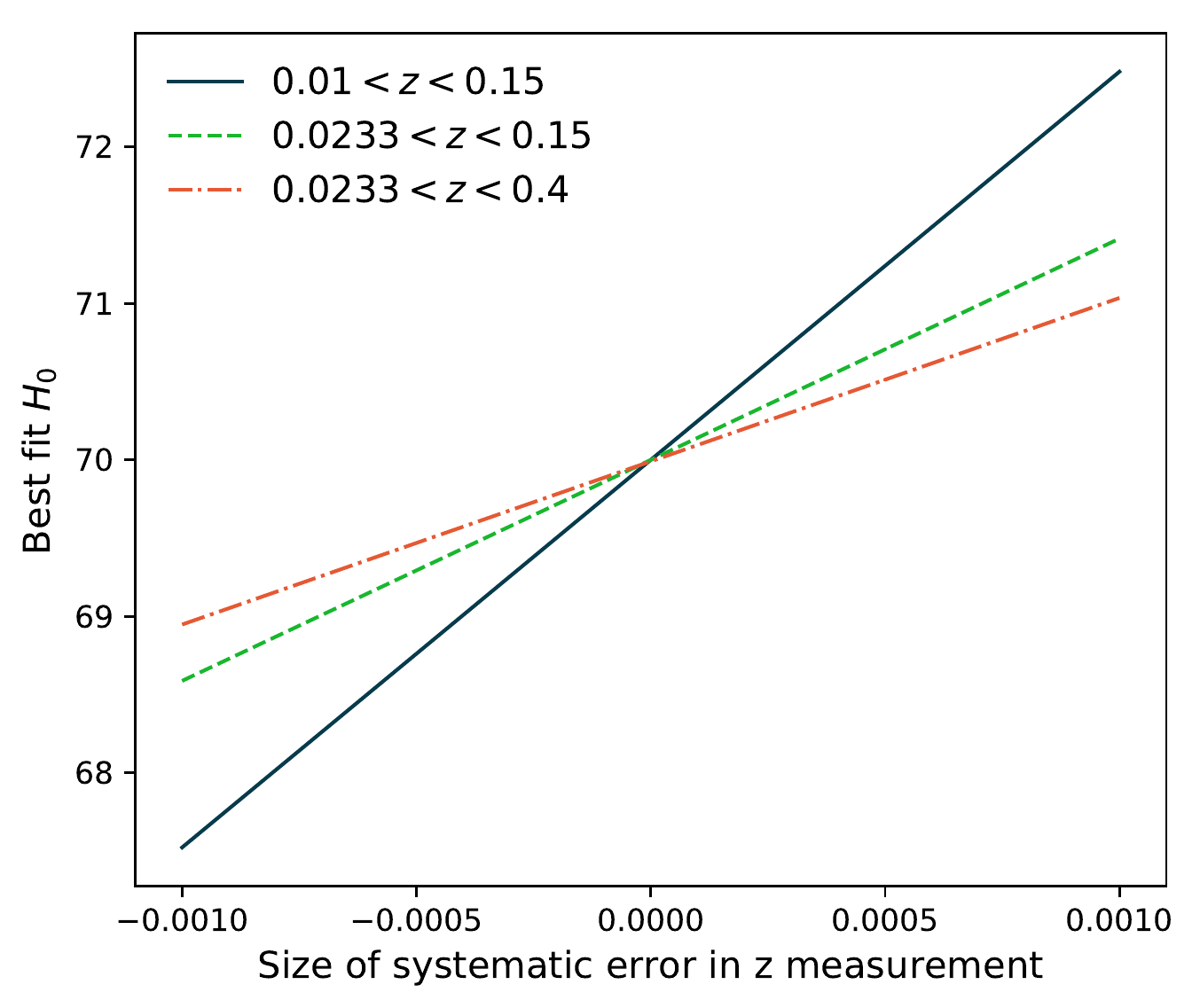}
    \caption{The value of $H_0$ (vertical axis) inferred in the presence of a systematic error in redshift (horizontal axis), for three different redshift ranges as shown in the legend.  The input $H_0=70$\kmsmpc.  This plot shows the bias for standard candles evenly distributed in log space across the redshift interval.  Different distributions across the redshift ranges (e.g.\ linear) would change the result, with more low-redshift (high-redshift) standard candles increasing (decreasing) the bias relative to that shown here.}
    \label{fig:H0error_vs_Deltaz}
\end{figure}

To summarise the impact of a redshift bias on the inferred value of $H_0$ we show in Fig.~\ref{fig:H0error_vs_Deltaz} the impact of a systematic error in $z$ on a standard candle measurement of $H_0$, as a function of the size of the systematic error.  In each case we evenly populate the redshift range in log space with $\sim200$ standard candles for $H_0=70$~\kmsmpc\ (as plotted in Fig.~\ref{fig:H0error_vs_z}).  We then add a redshift systematic $\Delta z$ of the size shown on the horizontal axis ($z_{\rm biased}=\bar{z}+\Delta z$) and measure $H_0$ from the biased data.   

Keeping any systematic error in redshift below $5\times10^{-4}$ and avoiding extremely low redshifts ($z_{\rm min}\gsim0.02$) limits the $H_0$ bias to less than 1~\kmsmpc.  However, we would like systematic biases to be at least an order of magnitude below the statistical uncertainty, and to keep the $H_0$ bias less than 0.1~\kmsmpc\ needs redshifts to be unbiased to better than $10^{-4}$.  Recall that statistical uncertainties on redshifts are often $\sim10^{-3}$.  

\section{How large could our redshift bias be?}\label{sect:large}
Therefore the question is, could there be a systematic offset in our measured redshifts at a level approximately an order of magnitude smaller than the uncertainties?  

Redshift errors can occur because of physical effects (e.g.\ peculiar velocities and gravitational redshifts), or measurement error (e.g. spectrograph wavelength calibration), or theoretical error (e.g. using the approximations in Sect.~\ref{sect:combiningz}).  In this section we assess how large several of these potential sources of error could be, and whether they would be systematic or random. 

\vspace{-5mm}\subsection{Local peculiar velocity corrections:}
Most cosmological calculations need to be done in the comoving (CMB) frame.  Therefore we need to remove the impact of our local peculiar velocities from the redshifts we measure using Eq.~\ref{eq:zs}.  In order of increasing importance:
\begin{itemize}
\item The rotation of the Earth ($<0.5$\kms; $z_{\rm p}^{\rm spin}\sim10^{-6}$), which is usually ignored as negligible, especially since we tend to point our telescopes as close to zenith as practical, which means we are looking perpendicular to the direction of motion;
\item The orbit of the Earth around the Sun ($30$\kms, $z_{\rm p}^{\rm Earth}\sim10^{-4}$), which is usually corrected by telescope software to the heliocentric redshift, that is the redshift that would be observed in the rest frame of our Sun;
\item The motion of the Sun relative to the CMB, $369.82\pm0.11$\kms\ \citep{planckI18}, corresponding to $z_{\rm p}^{\rm Sun}=(1.2343\pm0.0004)\times10^{-3}$, 
 which is not usually applied at the telescope and needs to be corrected by the user (and thus more likely to be susceptible to user error, e.g. using $v_{\rm tot}=cz_{\rm obs}$ or $\bar{z}=z-z_{\rm p}$ or a sign error, which would all be systematic errors).  
\end{itemize}
Note that all of the above peculiar velocities are direction dependent, and the redshifts mentioned are the maximum redshift correction that may be required.  

These three peculiar velocities are small enough that they can be converted to redshifts using $z_{\rm p}=v_{\rm p}/c$. 
For the Sun's motion with respect to the CMB that gives $z_{\rm p}^{\rm Sun}=1.2336\times10^{-3}$, while the full special relativistic formula\\ $1+z=\sqrt{\frac{1+v/c}{1-v/c}}$ gives $z_{\rm p}^{\rm Sun}=1.2343\times10^{-3}$, 
a difference that is now technically significant, given the high precision of the measurement -- but not important because it is still only an error in redshift of $\sim10^{-6}$.  Note that it is appropriate to apply the special relativistic velocity-redshift formula to peculiar velocities, but {\it not} to recession velocities.\vspace{-5mm}\footnote{Peculiar velocities are measured with respect to the local reference frame, and there is a local Minkowski frame at every infinitesimal point in the universe in which special relativity applies.  So the special relativistic velocity-to-redshift formula is appropriate to convert the peculiar velocity to the peculiar redshift (the redshift a comoving observer would see when they are coincident with the source).  However, recession velocities (the velocity that appears in the Hubble-Lema\^{i}tre law) is not a velocity in anyone's inertial frame and should {\it not} be converted using the special relativistic correction \citep{davis04}. 
At  $z\sim0.1$ using $v_{\rm tot}=cz_{\rm obs}$ overestimates the recession velocity by $+$700\kms, while using the special relativistic formula underestimates the recession velocity by $-800$\kms\ \citep[e.g.][Fig.~2.1]{davis03}.  
}   

Were someone to apply the heliocentric correction unnecessarily (not realising it had been done at the telescope), or in the wrong direction, it would result in a systematic redshift error of $\sim10^{-4}$.  If redshifts are not corrected to the CMB frame, or corrected in the wrong direction, it would result in a systematic redshift error of $\sim10^{-3}$.

\subsection{The equation used for redshift corrections:}\label{sect:zcorr}
Considering just the heliocentric to CMB contribution in Eq.~\ref{eq:zs} we can see that the correct way to convert from the Sun's reference frame to the cosmological one is,
\beq 1+\bar{z} = \frac{1+z_{\rm obs}}{1+z_{\rm p}^{\rm Sun}}. \eeq
If one uses the approximation $\bar{z}\approx z_{\rm obs}-z_{\rm p}^{\rm Sun}$ (or equivalently $c\bar{z}\approx cz_{\rm obs}-cz_{\rm p}^{\rm Sun}$) then
at $\bar{z}=0.1$ it leads to a redshift error of $1.2\times10^{-4}$, while at $\bar{z}=1.0$ the error increases to $1.2\times10^{-3}$ (see Fig.~\ref{fig:redshifts}).  These have usually been considered too small an offset to worry about, but Section~\ref{sect:zbias} shows that for modern cosmological measurements this approximation is no longer sufficient.  
Since the NED velocity conversion calculator recommends this approximation, it is likely that this error appears frequently in the literature.

\subsection{Overcorrection in the presence of a bulk flow}\label{sect:bulkflow}
Bulk flows are sourced on very large scales, and in $\Lambda$CDM the expected mean bulk flow remains large even at quite large distances.   A spherical volume with radius 100 $h^{-1}$Mpc ($z\sim0.033$) is still expected to have a bulk flow of 100 \kms\ \citep[see][Fig.~10]{scrimgeour16}.  Were we to correct to the CMB frame within that volume we would be overcorrecting the redshift on average \citep{mould00}.  

\begin{figure}
    \centering
    \includegraphics[width=84mm]{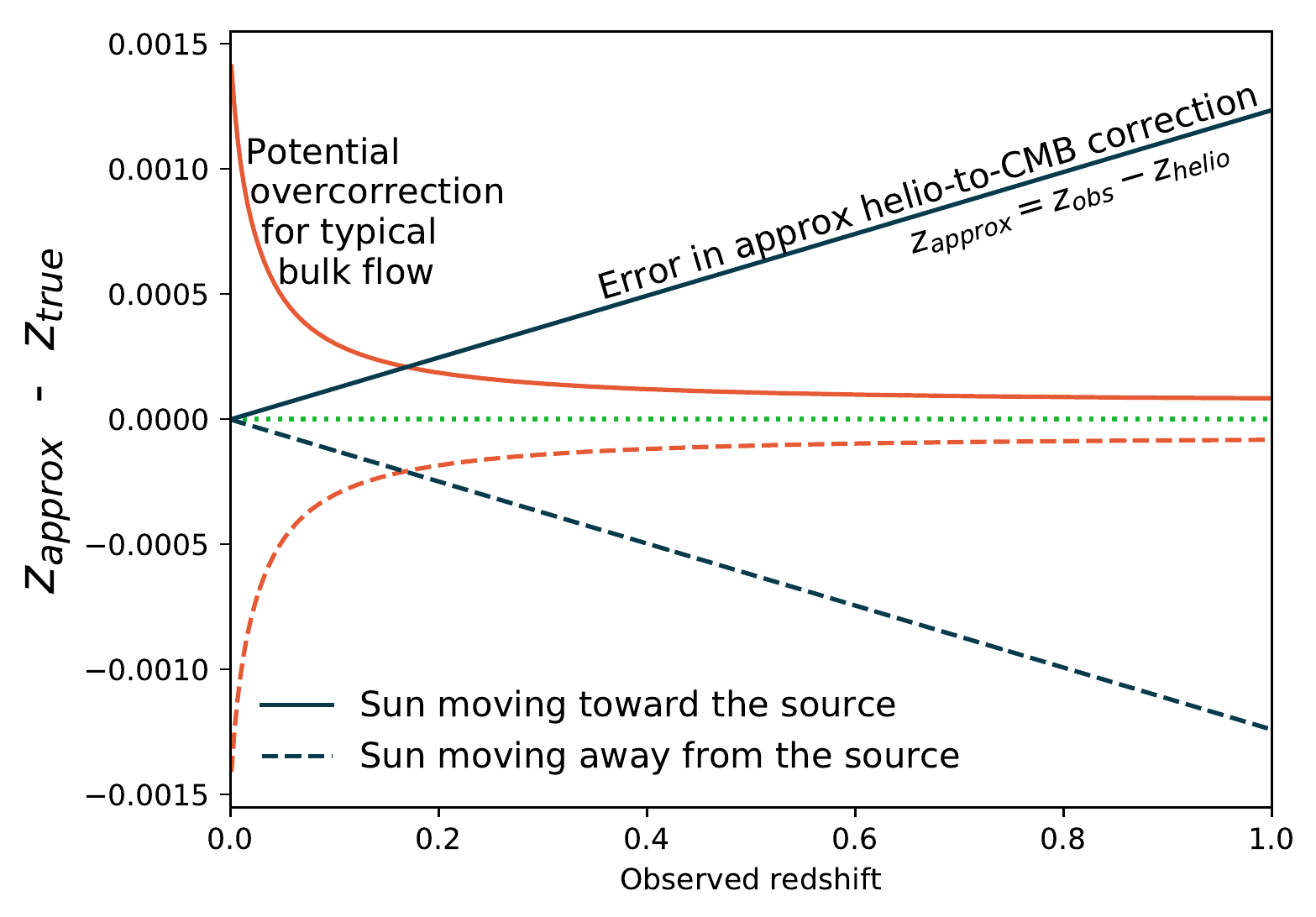}
    \caption{This plot shows two potential sources of systematic error.  Firstly, the error in redshift caused by using the approximation $\bar{z}_{\rm approx} = z_{\rm obs}-z_{\rm p}^{\rm Sun}$ to correct for our Sun's motion with respect to the CMB instead of the full formula $\bar{z}=\frac{1+z_{\rm obs}}{1+z_{\rm p}^{\rm Sun}}-1$ (Sect.~\ref{sect:zcorr}).  This is the maximum correction directly along the axis of the CMB dipole, i.e. $v_{\rm p}^{\rm Sun} = \pm369.82$\kms, 
corresponding to a redshift due to the motion of our Sun of $z_{\rm p}^{\rm Sun}=\pm0.0012343$.  Secondly, the potential bias in our redshifts if we correct for the heliocentric-to-CMB velocity, without taking into account that galaxies within quite a large volume around us are expected to share some of our bulk flow velocity (Sect.~\ref{sect:bulkflow}). }
    \label{fig:redshifts}
\end{figure}

As an example, imagine that we share the same peculiar velocity as a measured galaxy since we are in the same bulk flow.  The measured redshift is then the cosmological one, and correcting for our velocity relative to the CMB would imprint a redshift error of the size of our peculiar velocity.\footnote{Mathematically if our peculiar velocity in the direction of the observed galaxy causes a blueshift, \beq 1+z_{\rm p}^{\rm Sun}=\sqrt{(1-v_{\rm p}^{\rm Sun}/c)/(1+v_{\rm p}^{\rm Sun}/c)} \eeq and the observed galaxy's peculiar velocity $v_{\rm p}^{\rm source}=v_{\rm p}^{\rm Sun}$ directly away from us gives a redshift, \beq 1+z_{\rm p}^{\rm source}=\sqrt{(1+v_{\rm p}^{\rm source}/c)/(1-v_{\rm p}^{\rm source}/c)}\eeq then these terms perfectly cancel,
\beq 1+\bar{z}_{\rm true} = \frac{1+z_{\rm obs}}{(1+z_{\rm p}^{\rm source})(1+z_{\rm p}^{\rm Sun})} = 1+z_{\rm obs}. \eeq  
(Note we had to use the special relativistic formula because $z=v/c$ would not perfectly cancel.)
However, we do not usually have knowledge of the observed galaxy's peculiar velocity.  Therefore, we correct for our peculiar velocity only, which would give,
\beq 1+\bar{z}_{\rm estimate} = \frac{1+z_{\rm obs}}{1+z_{\rm p}^{\rm Sun}}.  \eeq 
}
In general the bias will be 
\beq \frac{1+\bar{z}_{\rm estimate}}{1+\bar{z}_{\rm true}}= (1+z_{\rm p}^{\rm Sun}),\eeq 

\beq \bar{z}_{\rm estimate}-\bar{z}_{\rm true} = \frac{z_{\rm p}^{\rm Sun}(1+z_{\rm obs})}{(1+z_{\rm p}^{\rm Sun})(1+z_{\rm p}^{\rm source})}=z_{\rm p}^{\rm Sun}(1+\bar{z}_{\rm true}).\eeq 
You can see from this last equation that another way to look at this is that we are simply not correcting for a systematic peculiar velocity (bulk flow).  
The magnitude of this error along the direction of motion for a typical bulk flow in $\Lambda$CDM is shown in Fig.~\ref{fig:redshifts}.
This is a directional dependent effect and should cancel out across the sky, but would be systematic for particular directions.

\subsection{Rest frame wavelength precision:} The rest frame wavelengths for spectral lines are usually programmed into optical spectrograph redshifting software with 0.01\AA\ precision \citep{hinton16,talbot18}.  An error in rest frame wavelength of 0.01\AA\ results in a redshift error of $\sim 5\times10^{-6}$ at $z\sim 1$ for \ion{O}{ii} (3727.09\AA), slightly less for higher wavelengths and at lower redshifts.  Rest frame wavelengths should therefore be sufficiently accurate for current cosmological applications.

One caveat is that if spectral lines are superimposed on a steep continuum, such as the calcium H and K lines often are, then the observed peak (or minimum) of the line will not precisely be at the line centre because of the continuum contribution.  For narrow lines this should be a small effect. 

Interestingly, for high precision redshift applications such as needed for tests of fine structure constant variation, knowledge of rest frame wavelengths has been a limiting factor -- both the wavelengths of the calibration lamps \citep{lovis07,redman14}, and the wavelengths of the spectral lines in the quasars being studied \citep{murphy14}.

\subsection{Air to vacuum conversion:} 

\begin{figure}
    \centering
    \includegraphics[width=84mm]{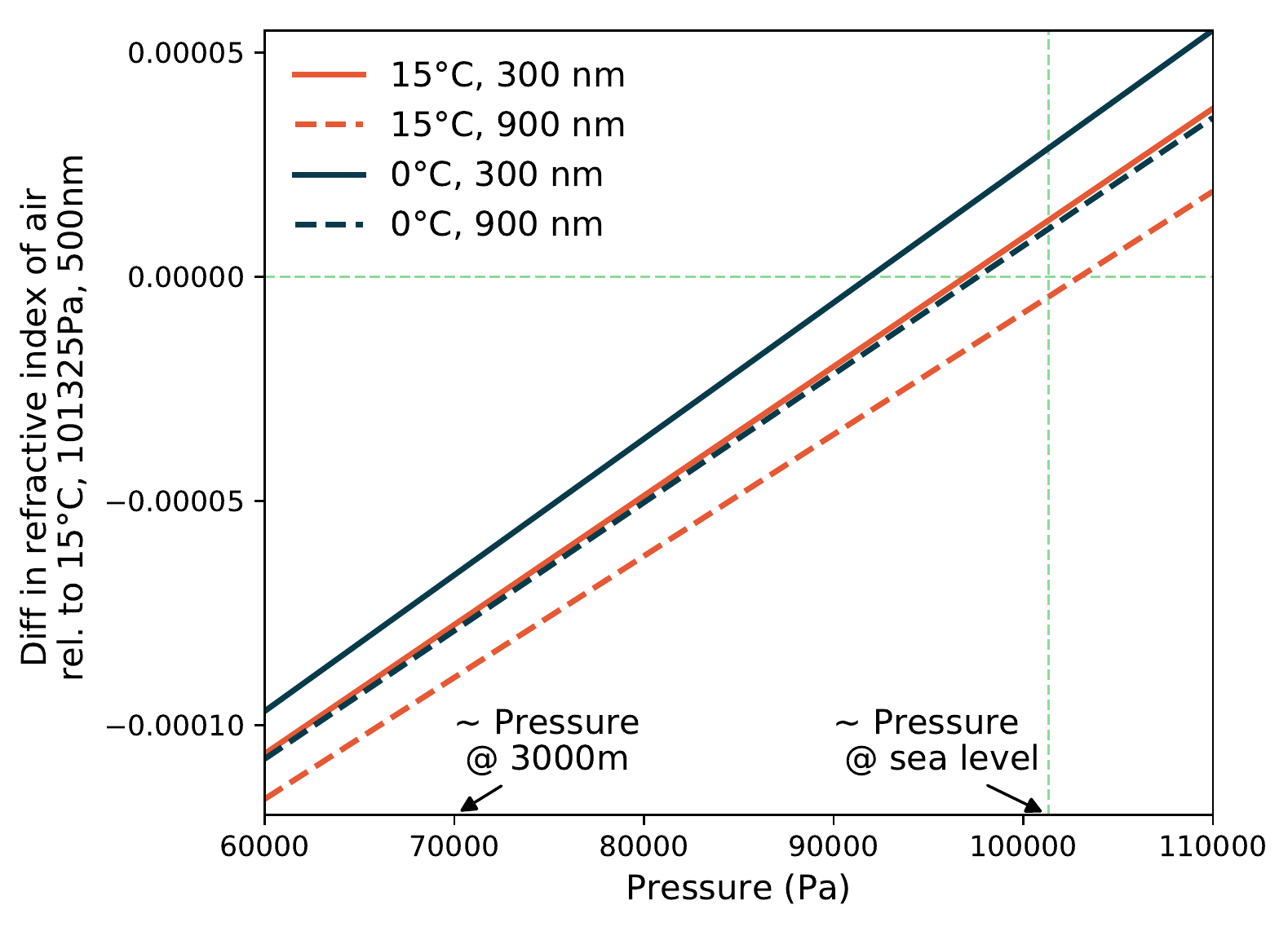}
    \caption{Difference in refractive index as a function of pressure for a wavelength of 300nm (solid) and 900nm (dashed), at 15$^\circ$C (red) and 0$^\circ$C (blue).  The difference shown on the vertical axis ($n-n_{\rm fid}$) is measured relative to a fiducial refractive index $n_{\rm fid}$ calculated at standard temperature and pressure (15$^\circ$C and 101kPa), and 500nm (dotted lines).   A typical atmospheric pressure at 3000m altitude is approximately 70kPa.  Humidity is kept at 0\% and CO$_2$ concentration 450ppm; reasonable variations in humidity or CO$_2$ concentration have at least an order of magnitude lower impact on refractive index than the pressures shown here.}
    \label{fig:n}
\end{figure}

One potential source of error is the conversion from air to vacuum wavelengths.  
The refractive index of air is $n_{\rm air}\equiv\lambda_{\rm vac}/\lambda_{\rm air}\sim1.00028$.  The air-to-vacuum wavelength conversion thus corresponds to a redshift of $\sim2.8\times10^{-4}$.
The refractive index of air depends on the temperature, pressure, and atmospheric composition, as well as the wavelength.\footnote{Air to vacuum wavelength conversions from SDSS \citep{talbot18} can be found at \url{https://classic.sdss.org/dr7/products/spectra/vacwavelength.html}, and take into account the differences at different wavelengths according to:
\beq n-1 = 2.735182\times10^{-4} + \frac{131.4182}{\lambda_{\rm vac}^2} + \frac{2.76249\times10^{8}}{\lambda_{\rm vac}^4}, \eeq 
which is equivalent (differs by $<10^{-7}$) to those found in \citet{ciddor96}:
\beq n-1 = \frac{0.05792105}{238.0185-\lambda^{-2}} + \frac{0.00167917}{57.362-\lambda^{-2}}.\eeq
The calculations done in this paper use the code provided by\\ 
\url{https://refractiveindex.info/?shelf=other\&book=air\&page=Ciddor}.}  It is $n_{\rm air}\sim1.00028$ at 500nm when conditions are 15$^\circ$C, 101325 Pa, 450ppm CO$_2$, and 0\% humidity.   At 3000m and 0$^\circ$C the air pressure is approximately 69000 Pa, and the index of refraction of air becomes $n_{\rm air}\sim1.00020$.  
Fig.~\ref{fig:n} shows refractive index vs pressure for two different temperatures and two wavelengths spanning the wavelength range of typical optical spectrographs \citep[adapting the code from][]{ciddor96}.

In practice it does not matter what the conditions are at your telescope, as long as the arc lamp exposures used for wavelength calibration are taken under the same conditions as the astronomical data.  So we would only get errors in the case of significant atmospheric changes between the time the wavelength calibration exposures are taken and the time of the observations. 

The air-to-vacuum conversions are used primarily to bring different spectral line libraries to a consistent baseline.  
As long as the conversion is done correctly so that redshifts are measured with respect to the correct wavelengths, and line libraries are accurate to at least six significant figures, systematic errors should be limited to $\Delta z < 10^{-6}$.

\subsection{Spectrograph wavelength calibration:}  It is possible to calibrate spectrographs extremely well (absolute redshift accuracy better than $10^{-7}$) when required for purposes such as exoplanet discovery using, for example, superimposed spectra from iodine cells or thorium argon lamps \citep{lovis07}, or laser frequency combs \citep{steinmetz08}.  However, \citet{whitmore15} use a `supercalibration' from asteroids and solar twins to show that even with these techniques systematic errors in wavelength calibration on the order of $10^{-6}$ are common. 

On lower resolution instruments without these technologies wavelength calibration can still be done very well.   SDSS state their  wavelength calibration is better than 5km\,s$^{-1}$, corresponding to a redshift error of 2$\times10^{-5}$.  Moreover, to be pathological it would have to be systematic, not random.   Therefore it is unlikely to cause a systematic large enough to account for the $H_0$ tension.  However, we do see occasional failures in wavelength calibration, for example, the wavelength calibration can be systematically worse at the edges of the wavelength ranges compared to the centre.  If this was to occur then all galaxies in a particular redshift range -- for example with \ion{O}{ii} at the blue end of the spectrograph --  would be systematically offset from galaxies that used \ion{O}{ii} at the centre of the spectrograph wavelength range.   

\begin{figure*}
    \centering
    \includegraphics[width=175mm]{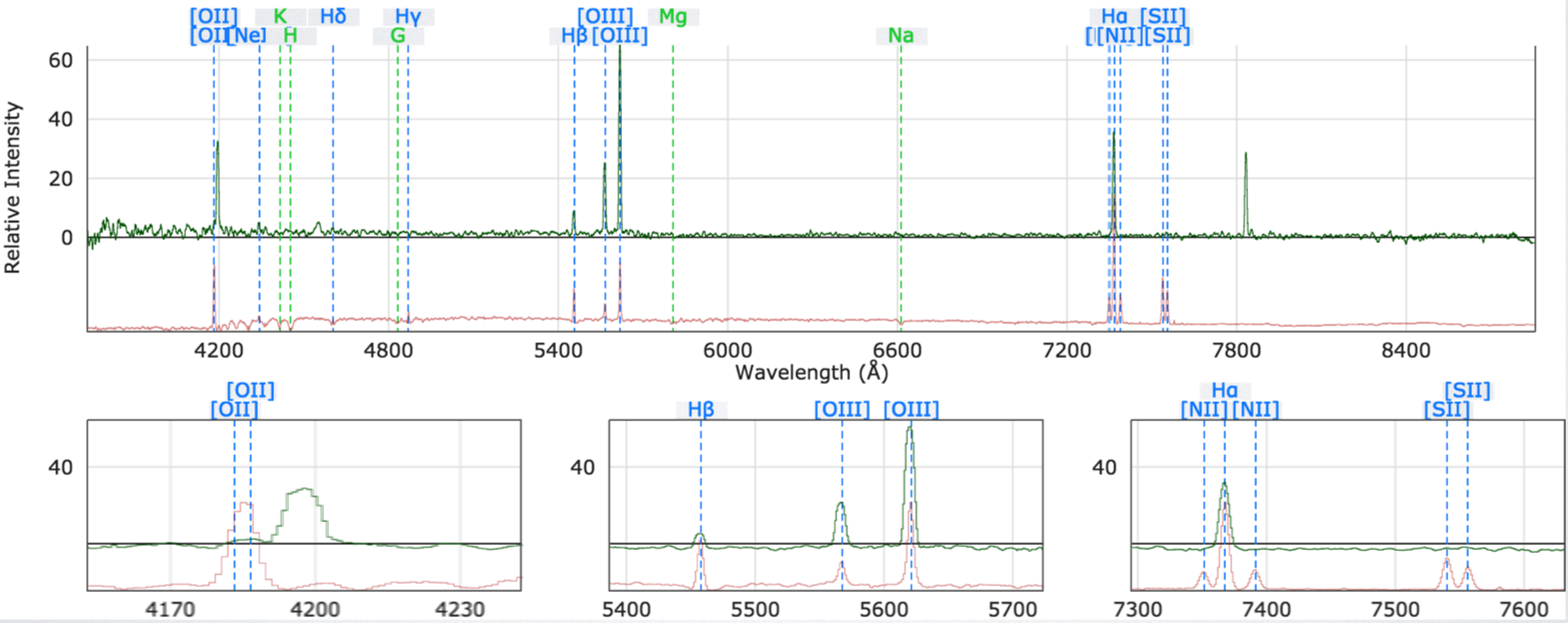}
    \caption{AAT spectrum showing an example of wavelength calibration failure.  The green line is the measured spectrum, the red line is a template spectrum from the database, the blue vertical lines are the expected positions of emission features.  Clearly the high-wavelength lines are offset from the expected position of the OII doublet.  While rare, if such failures happen systematically in one direction it can shift the mean of the best fitting redshifts.  It may also affect low- and high-redshift galaxies differently as different lines appear at different wavelengths.  This plot actually has two spectra plotted on top of one another, both in green lines, separated by a redshift of $10^{-4}$ to show on data what such a small shift looks like.  If you are finding it difficult to see, that is because the shift is generally smaller than the width of the line on this display, it is most noticeable  in the lower left panel showing the \ion{O}{ii} line slightly offset (the difference between the two green lines, not the difference between the green observed spectrum and red template spectrum, which is $\Delta z \sim 3\times10^{-3}$).}
    \label{fig:spectrum}
\end{figure*}

An example of a pathological wavelength calibration failure can be seen in Fig.~\ref{fig:spectrum}, which shows a spectrum from the Anglo-Australian Telescope using the Marz spectral viewer and redshifting tool \citep{hinton16}, for which the wavelength solution at high-wavelengths (H$\alpha$, H$\beta$, and  \ion{O}{iii}) is clearly incompatible with that at low wavelengths (\ion{O}{ii}).  The red template spectrum aligns with the expected line positions (blue dashes) whereas the green observed spectrum cannot simultaneously fit the \ion{O}{ii} line and the higher wavelength lines.  On this plot we also show how small a $10^{-4}$ shift appears in spectral data -- we plot two green spectra separated by $10^{-4}$, which are almost indistinguishable because the shift is approximately the width of the line. 

\subsection{Spectrograph resolution and redshift measurement:}
The SDSS has a spectral resolution of $R=\lambda/\Delta\lambda = 1500$ at 3800\AA\ and 2500 at 9000\AA, \url{http://www.sdss3.org/dr9/spectro/spectro_basics.php} while OzDES has $R=1800$ \citep{childress17}.  These correspond to redshift resolution of $4\times10^{-4}$ to $7\times10^{-4}$.  The redshift precision quoted by these surveys is thus on the order of the pixel size.   

For example, the OzDES redshift survey gives an uncertainty of $4\times10^{-4}$ for galaxy redshifts and $1.5\times 10^{-3}$ for AGN redshifts \citep{yuan15}.  These were calculated both through internal dispersion within the catalogue, and by comparison with external catalogues.  This magnitude of uncertainty was again confirmed in \citet{childress17} who measured an rms dispersion in redshift recovery of $\sigma_z=4.2\times10^{-4}$ between initial science-quality redshift measurements (qop=3), and subsequent higher signal-to-noise measurements of the same source (qop=4).
The uncertainties are calculated as weighted pair-wise redshift differences, $\sigma_z=\Delta z/(1 + z)$, which means that at $z\sim 1$ the uncertainties are not $4\times10^{-4}$ but $8\times10^{-4}$ after multiplying by $(1+z)$.  

Note, however, there is a risk when using internal consistency to estimate redshift uncertainties, or even comparison with other survyes, because there is the possibility of hidden systematics that would not be picked up by such tests.  Absolute wavelength calibration also needs to be monitored, for example by using sky lines to test the wavelength solution.  The sky lines have gone through the same optical path as the astrophysical source (at least from the top of the telescope to the spectrograph) and should not suffer from velocity shifts due to bulk motions.

\vspace{-3mm}
\subsection{Density fluctuations:} If we live in a local over- or under-density then the same gravitational redshift will affect all photons that fall into our local potential well or climb our local potential hill.  The expected magnitude of such a shift in a standard $\Lambda$CDM model given expected density fluctuations is on the order of $\sim10^{-5}$ \citep[see][Figs~1-3]{wojtak15}.  \citet{kenworthy19} summarise the recent evidence for a local overdensity or void from the supernova data, concluding that there is no evidence for a significant local void, and any local density fluctuation on scales larger than 69 Mpc $h^{-1}$ must have a density contrast  $\delta=(\rho-\bar{\rho})/\bar{\rho}$ within  $|\delta| < 27\%$.

\vspace{-4mm}
\subsection{Internal velocities:}
Galaxies rotate, and have inflows and outflows.  If redshifts are gauged from the edge of a spiral galaxy, rather than the centre, then they may be subject to peculiar velocity shifts on the order of $\sim300$\kms, corresponding to $\Delta z\sim10^{-3}$.  There is a bias toward discovering supernovae away from the centre of a galaxy, because of less dust obscuration and less chance of it being mistaken for an AGN.  The redshifts of some supernova hosts will have been obtained by galaxy light near the location of the supernova, and therefore may include the peculiar velocity of the rotation instead of the systemic redshift of the galaxy.  This would affect nearby galaxies more than distant ones, because they are resolved and only part of the galaxy will fall within the slit or fibre of the spectrograph.  However, it is an error that should be primarily random, as as many should be receding from us as approaching us.

If a galaxy has an outflow, then we will see the part of it that is coming toward us more than that going away, and the systemic velocity of the galaxy may appear blueshifted.  This effect would be systematic. 

Interestingly, even in the absence of an outflow, we would be slightly biased by the internal velocity dispersion of galaxies because stars and gas that are approaching us appear brighter due to relativistic beaming than those moving away, even if the velocity distribution is symmetric.  Assuming flux is proportional to temperature, and a Gaussian velocity distribution, galaxies will be systemically blueshifted by $\Delta z \sim (\sigma/c)^2$.  
However, it is a small effect and only contributes $10^{-6}$ even for giant ellipticals \citep{kochanek19}. 


\section{Conclusion}

In the era of modern cosmology where precision measurements are the norm, we have shown it is timely to revisit the accuracy with which we measure redshifts.  Small systematic redshift errors matter when measuring cosmological parameters to 1\% precision.  The errors do not have to be perfectly systematic -- if the mean redshift deviates from the true mean by the amounts we discuss here, then that is effectively a systematic error and would have a similar effect.  Additive systematic errors at low redshifts have a higher impact on cosmological results than the same error at high redshifts.  

Possibly most importantly, there are common theoretical approximations that have been permissible in the past, but are no longer appropriate given the quality and depth of our data.  We advocate for working with measured quantities (redshifts) as much as possible without converting to velocities.  Using $(1+z_{\rm obs})=(1+\bar{z})(1+z_{i})$ with as many $i$ terms as necessary to account for peculiar velocities, gravitational redshifts, etc... is computationally as simple as adding redshifts, and ensures accuracy.  One should never use $v_{\rm tot}=cz_{\rm obs}$, but instead use Eq.~\ref{eq:correctvz} or its approximation Eq.~\ref{eq:vzHD}.  Using two different redshifts in luminosity and angular diameter distance is preferable, but is a small effect and using only cosmological redshift for all terms in $D_L(\bar{z},z_{\rm obs})=(1+z_{\rm obs})\tilde{D}(\bar{z})$ remains a good approximation. 

The most likely observational issues to affect redshifts are occasional wavelength calibration errors that impact the mean redshift.   The most likely physical effects that impact redshifts are peculiar velocity corrections (including overcorrecting for our motion when nearby galaxies ($z\lsim 0.04$) systematically share some of that motion), internal velocities of galaxies, and outflows. 

We conclude that before turning to exotic theory to explain tensions in cosmological parameters, it remains worthwhile to carefully assess whether any systematic errors in redshifts are present.

\section*{Acknowledgements}

We thank Chris Blake, Florian Beutler, and Michael Murphy for useful discussions.  This research was supported by the Australian Government through the Australian Research Council's Laureate Fellowship funding scheme (project FL180100168).




\bibliographystyle{mnras}
\bibliography{h0} 



\appendix

\section{Comparison to literature nomenclature}\label{app:riess}

\begin{figure*}
    \centering
    \includegraphics[width=80mm]{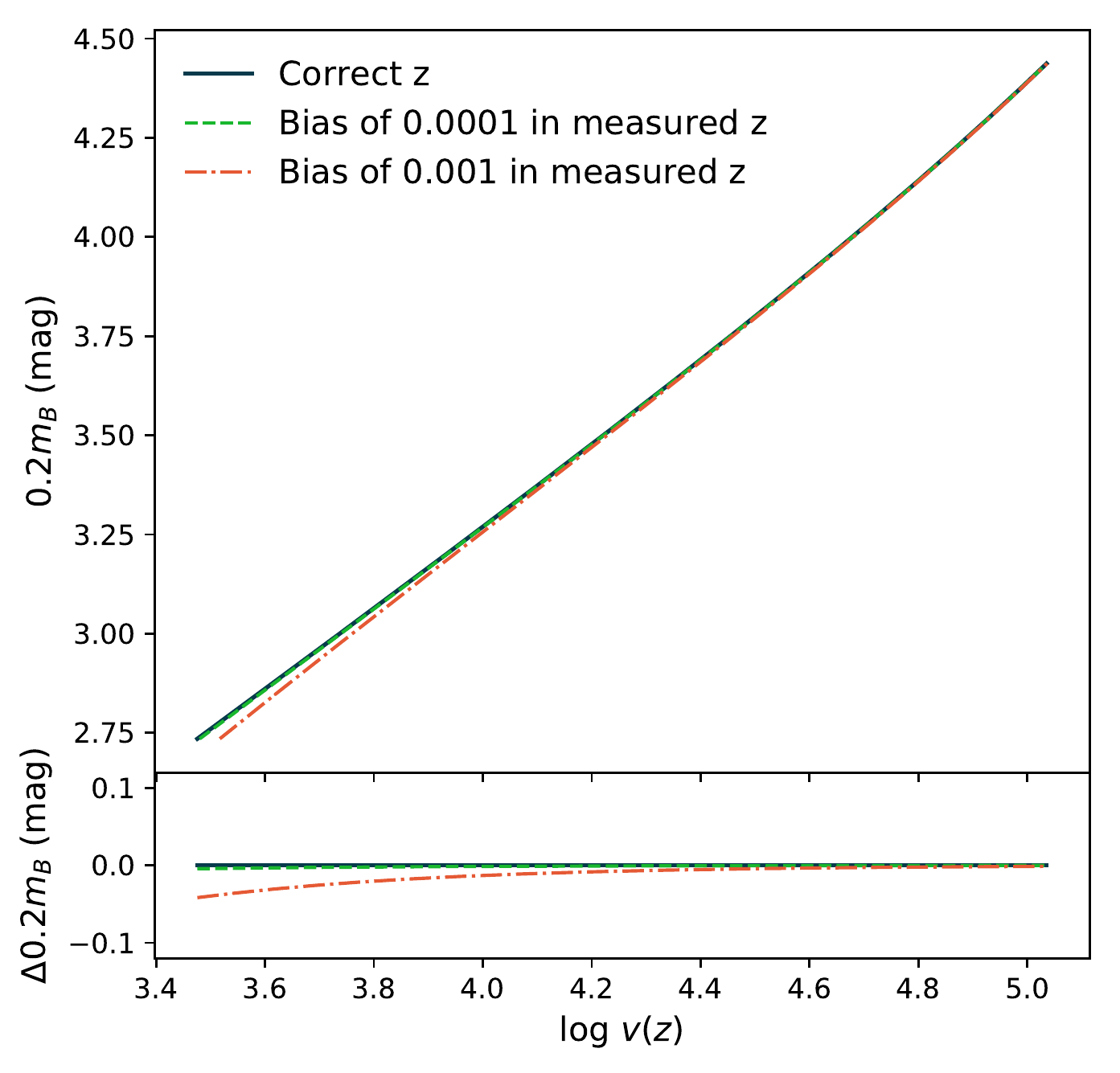}
    \includegraphics[width=90mm]{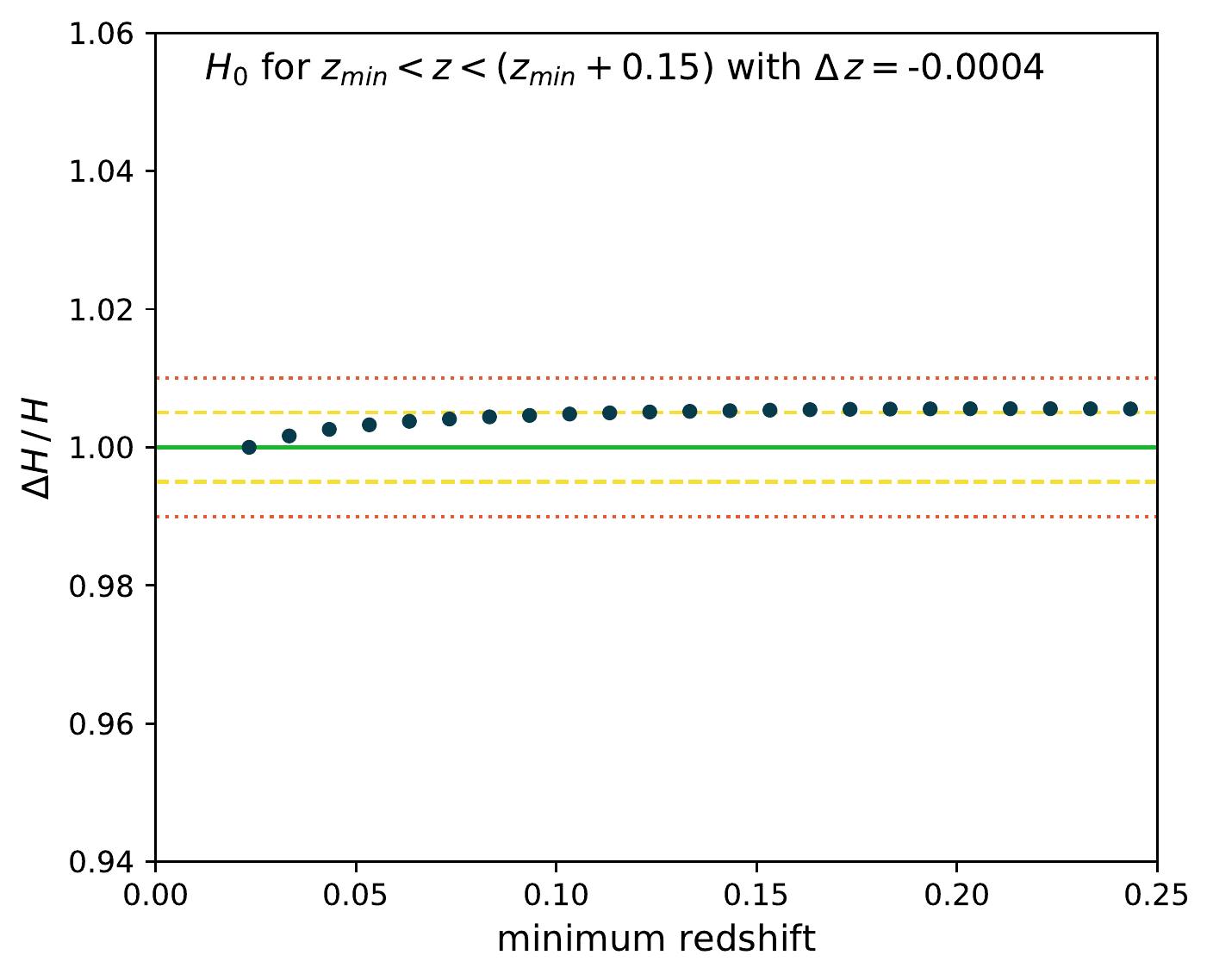}
    \caption{Left: What a redshift error would look like on a plot of magnitude vs velocity \citep[Fig.~8 of][]{riess16}.  The extreme redshift systematic of $10^{-3}$ would likely be noticeable if it was perfectly systematic, but a systematic of $10^{-4}$ would not be identifiable on this plot.  Right: Reproduction of the form of \citet{riess16} Fig.~12, implementing a redshift bias of $-4\times10^{-4}$.  We interpreted the figure to be $(H_0(z_{\rm min}) - H_0(0.0233))/H_0(0.0233)$.  If we have interpreted the direction of the $\Delta H$ correctly, then this plot actually shows it is unlikely for a redshift bias to have caused an {\it overestimate} of $H_0$, because this shift needs a negative redshift bias, but to overestimate $H_0$ you would need a positive redshift bias.  These axis ranges mimic \citet{riess16} exactly, but the plot on the right is not a perfect reproduction of their data because we have used a smooth redshift distribution instead of the actual redshift distribution of their sample.}
    \label{fig:02mv}
    \label{fig:lowerzlim}
\end{figure*}

$H_0=v/D$, may not immediately look like $\log_{10}H_0=(M+5a_x+25)/5$, which is the equation used to derive $H_0$ in \citet{riess16} and similar papers.  However, a simple transform takes one to the other.
First recall that the definition of distance modulus is $\mu=m-M=5\log_{10}D_L({\rm Mpc})+25$, so $\log_{10}D_L = \frac{m-M-25}{5}=0.2m-\frac{M+25}{5}$.  
In a flat universe $D=\tilde{D}$ so:
\begin{eqnarray}
H_0&=&\frac{v}{D} \\
      &=&\frac{v(1+z)}{D_L} \\
\log_{10} H_0&=&\log_{10} [v(1+z)]-\log_{10} D_L \\
                      &=&\log_{10} [v(1+z)]-0.2m+\frac{M+25}{5} \\
                      &=&\frac{5a_x+M+25}{5} 
\end{eqnarray}
where $a_x\equiv \log_{10} [v(1+z)]-0.2m$.
Adding curvature corrections requires us to write $D_L\equiv\tilde{D}(1+z)=D\frac{S_k(\chi)}{\chi}(1+z)$, so we just need one extra term in $a_x\equiv \log_{10} [v(1+z)]-0.2m+\log_{10}[S_k(\chi)/\chi]$.  Since an approximation is used for $a_x$ in any case (see Eq.~\ref{eq:vzHD}), and the curvature term is small, this is a negligible source of error. 

To examine whether a redshift bias would have been noticed in existing standard candle data we have reproduced in Fig.~\ref{fig:02mv} the standard plot of $0.2 m_B$ vs the $v(z)$ approximation shown in \citet{riess16} Fig.~8.    We conclude that it would not be possible to notice a $10^{-4}$ systematic on this plot, but a systematic of  $10^{-3}$ might be identifiable.  The reason that such a small shift in redshift can have such a large impact on $H_0$ is because the value of $H_0$ essentially comes from deriving the $y$-intercept of this curve, which is a long way off the horizontal scale of this plot -- therefore a small error in slope generates a large error in intercept. 

Finally, we look at the impact of changing the lower redshift limit on the derivation of $H_0$, and consider what magnitude of redshift shift would be needed to reproduce the pattern seen in \citet{riess16} Fig.~12.  This is shown in Fig.~\ref{fig:lowerzlim}, where a redshift shift of $-4\times10^{-4}$ matches the size of the offset with redshift, although not the functional form.


\bsp	
\label{lastpage}
\end{document}